\documentclass{aa}
\usepackage{amsbsy}
\usepackage{amssymb}
\usepackage[dvips]{graphicx}

\usepackage{txfonts}
\usepackage{natbib}

\begin{document}

\title
{
  Gamma-ray burst contributions
  to constraining the evolution of dark energy
}

\author
{
  Shi Qi\inst{1,4}
  \and
  Fa-Yin Wang\inst{2}
  \and
  Tan Lu\inst{3,4}
}

\institute
{
  Department of Physics, Nanjing University, Nanjing 210093, China \\
  \email{qishi11@gmail.com}
  \and
  Department of Astronomy, Nanjing University, Nanjing 210093, China \\
  \email{fayinwang@nju.edu.cn}
  \and
  Purple Mountain Observatory, Chinese Academy of Sciences, Nanjing
  210008, China \\
  \email{t.lu@pmo.ac.cn}
  \and
  Joint Center for Particle, Nuclear Physics and Cosmology, Nanjing
  University -- Purple Mountain Observatory, Nanjing  210093, China
}

\abstract
{}
{
  We explore the gamma-ray bursts' (GRBs') contributions in
  constraining the dark energy
  equation of state (EOS)
  at high ($1.8 < z < 7$) and at middle redshifts
  ($0.5 < z < 1.8$) and estimate how many GRBs are needed to get
  substantial constraints at high redshifts.
}
{
  We estimate the constraints with mock GRBs and
  mock type Ia supernovae (SNe Ia) for comparisons.
  When constraining the dark energy EOS in a certain redshift range,
  we allow the dark energy EOS parameter to vary only in that redshift
  bin and fix EOS parameters elsewhere to $-1$.
}
{
  We find that it is difficult to constrain the dark energy EOS
  beyond the redshifts of SNe Ia with GRBs unless some new luminosity
  relations for GRBs with smaller scatters are discovered.
  However, at middle redshifts, GRBs have comparable contributions
  with SNe Ia in constraining the dark energy EOS.
}
{}


\maketitle

\section{Introduction}

Since the discovery of the accelerating expansion of the
universe~\citep{Riess:1998cb, Perlmutter:1998np}, a lot of work has
been done on constraining the behavior of dark energy. In contrast to
simple parametric forms for dark energy equation of state (EOS),
such as $w(z) = w_0+w'z$~\citep{Cooray:1999da} and
$w(z)=w_0+w_az/(1+z)$~\citep{Chevallier:2000qy, Linder:2002et},
a nearly model-independent approach was introduced
by~\citet{Huterer:2004ch} to estimate the evolution of
dark energy independently, and was adopted in analyses using
type Ia supernova (SN Ia)
data~\citep{Riess:2006fw, Sullivan:2007pd, Sullivan:2007tx}.
\citet{Qi:2008zk} extended the approach to
gamma-ray burst (GRB) luminosity data~\citep{Schaefer:2006pa},
for which higher redshifts are accessible compared to SNe Ia.
Though more
stringent results have been obtained by this, the dark energy EOS at
high redshifts,
where we only have GRBs,
is still totally unconstrained, except that it is most likely
negative. This is primarily due to matter dominating dark energy
at high redshifts and to dark energy becoming less important for
determining the cosmic expansion; therefore, constraining it becomes
more difficult. In this work we explore both how many GRBs are needed
to
get substantial constraints on dark energy EOS at high redshifts
beyond those of SNe Ia and where
the GRBs' contributions lie for a foreseeable number of GRBs.

\section{Methodology}

For the uncorrelated estimates of dark energy EOS, we first separate
the redshifts into several bins and assume a constant dark energy EOS
for each bin. Then we generate a Markov chain through the Monte-Carlo
algorithm according to the likelihood function. The covariance matrix
of the dark energy EOS parameters is calculated based on the Markov
chain, and a transformation derived from the covariance matrix
to decorrelate the EOS parameters. The evolution of dark energy is
finally estimated out of the uncorrelated EOS parameters
(see~\citet{Huterer:2004ch} for details).

The process of generating Markov chains in these procedures is
very time-consuming,
especially when there are considerable observational/mock standard
candles. We see that we need a large number of GRBs to get
substantial constraints on dark energy EOS at high redshifts.
Instead of using the uncorrelated approach directly, we adopt an
alternative way to estimate this number.
When estimating the constraints imposed on dark energy
EOS in a certain redshift range, we allow the EOS parameter to vary
only in that redshift bin and fix the EOS parameters elsewhere, say,
to $-1$.
This is compatible with the uncorrelated approach to an approximate
estimate, because the uncorrelated EOS parameters
defined in~\citet{Huterer:2004ch} are themselves localized well in
redshift bins. Of course, the fixation of EOS at some redshifts will
cause underestimates of the errors on the EOS at other redshifts, so
the number we have estimated here can be viewed as a lower limit.

We determine the likelihood for EOS parameters from the $\chi^2$
statistic,
\begin{equation}
  \label{eq:chi2_statistic}
  \chi_n^2 = \sum_{i = 1}^{n}
  \frac
  {
    \left[
      \mu_{t, i}(\theta) - \mu_{o, i}
    \right]^2
  }
  {
    \sigma_i^2
  }
  ,
\end{equation}
where $\mu_o$ is the observational/mock distance modulus of the
standard candles with standard deviation $\sigma$,
$\mu_t(\theta)$ is the theoretically predicted
distance modulus based on a cosmological model with parameter set
$\theta$, and $n$ is the number of the standard candles.
We constrain dark energy EOS parameters with mock data, whose
generation involves random numbers (see the details below). To reduce
the impact from the fluctuation of the mock
data themselves, we generate many more standard candles than
needed. For example, if we want to see the constraints imposed by $n$
samples of standard candles, we actually generate $N$ mock samples
with $N \gg n$ and calculate $\chi_N^2$. Then $\chi_n^2$ is given by
\begin{equation}
  \label{eq:mean_chi2}
  \chi_n^2 = \frac{n}{N}\chi_N^2
  .
\end{equation}
Throughout this paper, we choose $N = 1.0\mathrm{E}5$.

The fiducial cosmological model we used to generate mock data is the
flat $\Lambda$CDM model with $\Omega_m = 0.279$ and $H_0 = 70.1
\mathrm{km/s/Mpc}$ (see~\citet{Komatsu:2008hk}). For SNe Ia, we
approximate
the intrinsic noise in distance moduli as a Gaussian scatter with a
dispersion of $0.1$ magnitudes and the observational errors as
$0.23$ magnitudes, which are approximately the averages of the
observational errors of the presently available SNe
Ia~\citep{Riess:2006fw, WoodVasey:2007jb, Davis:2007na}. The total
error in our generated distance moduli for SN Ia is therefore
$\sqrt{0.1^2+0.23^2}$ magnitudes. We assume
an uniform distribution for SNe Ia along the redshifts. For GRBs,
instead of generating mock data about the five luminosity
relations (see~\citet{Schaefer:2006pa}), we directly generate
distance moduli like we do for SNe Ia for simplicity. The intrinsic
scatter is set to be $0.65$ magnitudes, which is approximately the
average of the
errors of the GRBs' average distance moduli presented
in~\citet{Schaefer:2006pa}, and we ignore the measurement
uncertainties, which are less than the intrinsic scatter.
We consider two kinds of distributions for GRBs
in the redshift bin $1.8 < z < 7$.
One is the uniform distribution, the other a very rough
approximation to the distribution presented by Fig.~2
in~\citet{Bromm:2005ep}, i.e. $P(z) \propto \exp (-z/7)$.
We will see that our results are independent of the GRB
distributions.

\section{Results}

\begin{figure}[htbp]
  \centering
  \includegraphics[width = 0.5 \textwidth]{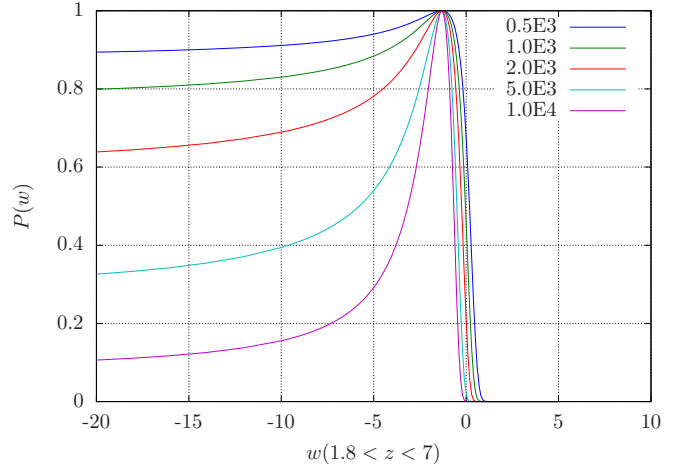}
  \caption
  {
    Constraints from GRBs uniformly distributed in the redshift bin
    $1.8 < z < 7$ on the dark energy EOS parameter $w (1.8 < z
    <7)$. The different lines stand for different number of GRBs. The
    probabilities are normalized to be $1$ at the maxima.
  }
  \label{fig:pw_GRB_hz_uni}
\end{figure}

\begin{figure}[htbp]
  \centering
  \includegraphics[width = 0.5 \textwidth]{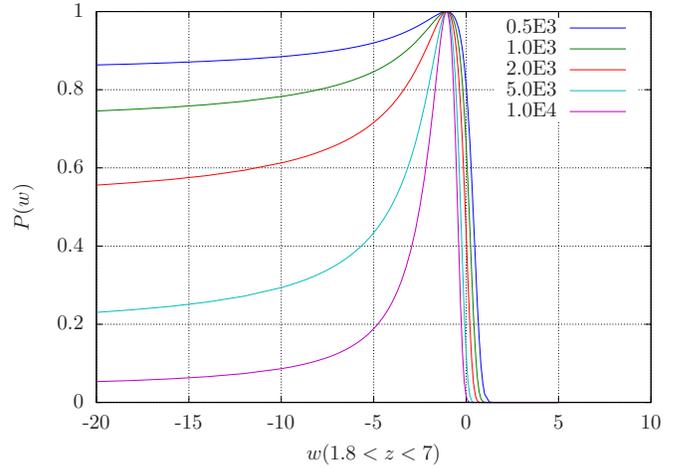}
  \caption
  {
    Constraints from GRBs distributed in the redshift bin
    $1.8 < z < 7$ according to $P(z) \propto \exp(-z/7)$ on the dark
    energy EOS parameter $w(1.8 < z < 7)$.
  }
  \label{fig:pw_GRB_hz_bl}
\end{figure}

Figures~\ref{fig:pw_GRB_hz_uni} and~\ref{fig:pw_GRB_hz_bl} show our
results for the constraints from GRBs distributed in the redshift bin
$1.8 < z < 7$ on the dark energy EOS parameter $w (1.8 < z < 7)$. We
can see that, for a few hundred GRBs, the constraints are only the
steep drop
at about zero that is seen in the probability function of the EOS
parameter $P(w)$. This
is consistent with the results in~\citet{Qi:2008zk}. Only when we have
more than about $5000$ GRBs can we begin to get concrete constraints
on the EOS parameter. The GRBs' distributions have little impact on
the conclusion; i.e., it is difficult to constrain the dark energy EOS
parameters beyond the redshifts of SNe Ia with GRBs unless some new
luminosity relations for GRBs with smaller scatters are discovered.

\begin{figure}[htbp]
  \centering
  \includegraphics[width = 0.5 \textwidth]{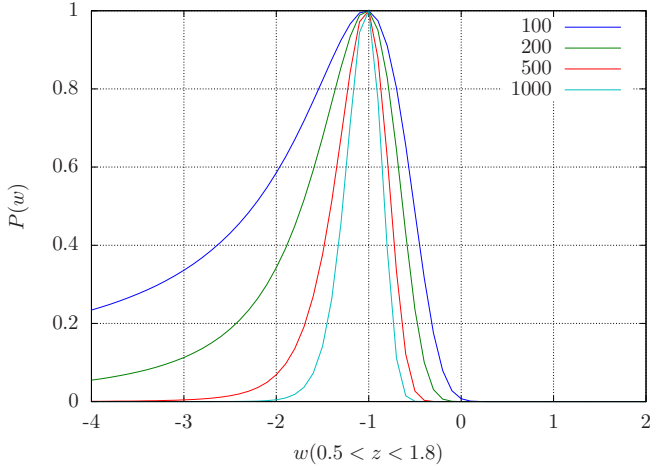}
  \caption
  {
    Constraints from GRBs uniformly distributed in the redshift bin
    $1.8 < z < 7$ on the dark energy EOS parameter
    $w (0.5 < z < 1.8)$.
  }
  \label{fig:pw_GRB_mz_uni}
\end{figure}

\begin{figure}[htbp]
  \centering
  \includegraphics[width = 0.5 \textwidth]{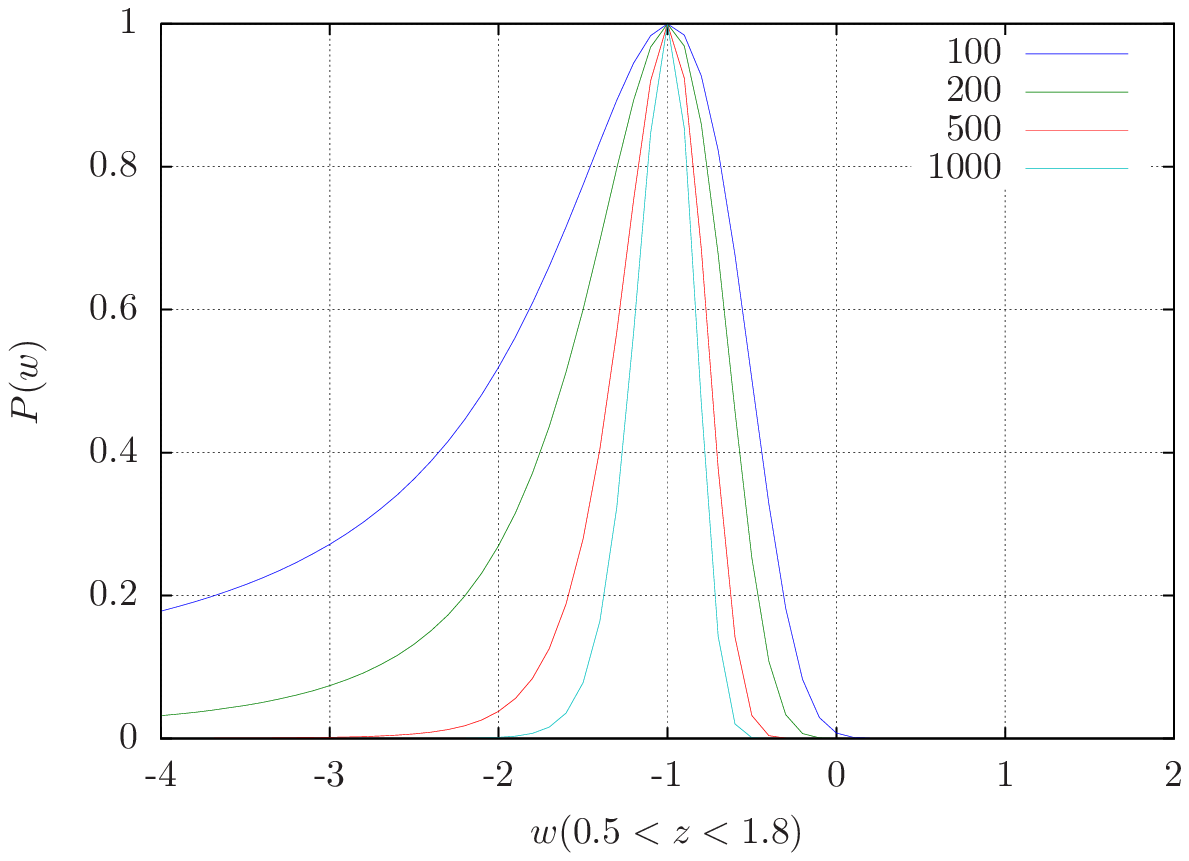}
  \caption
  {
    Constraints from GRBs distributed in the redshift bin
    $1.8 < z <7$ according to $P(z) \propto \exp(-z/7)$ on the dark
    energy EOS parameter $w (0.5 < z < 1.8)$.
  }
  \label{fig:pw_GRB_mz_bl}
\end{figure}

\begin{figure}[htbp]
  \centering
  \includegraphics[width = 0.5 \textwidth]{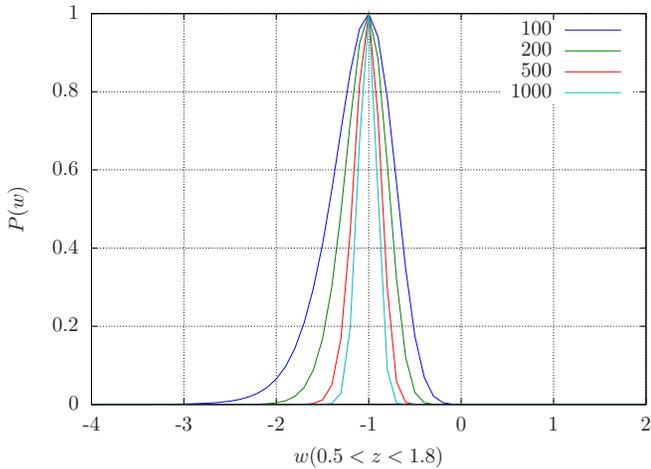}
  \caption
  {
    Constraints from SNe Ia uniformly distributed in the redshift bin
    $0.5 < z < 1.8$ on the dark energy EOS parameter
    $w (0.5 < z < 1.8)$.
  }
  \label{fig:pw_SNIa_mz_uni}
\end{figure}

However, this does not mean that high-redshift GRBs
contribute little to
constraining the dark energy EOS parameters. It has been demonstrated
in~\citet{Qi:2008zk} that, even with the presently available 69
GRBs~\citep{Schaefer:2006pa}, the constraints could be improved
significantly at redshifts $0.5 \lesssim z \lesssim 1.8$.
Part of the improvement stems from GRBs beyond redshift $1.8$.
Because the luminosity distances of standard candles depend on the
behavior of the dark energy through an integration over the redshift,
high-redshift GRBs put constraints on dark energy at lower redshifts,
where dark energy is important in determining the cosmic expansion.
And since there are few GRBs at low redshifts, the contributions from
GRBs would lie primarily in the middle redshifts.
In Figs.~\ref{fig:pw_GRB_mz_uni} and~\ref{fig:pw_GRB_mz_bl} we
explicitly show the constraints from GRBs distributed in the redshift
bin $1.8 < z < 7$ on the dark energy EOS parameter
$w (0.5 < z < 1.8)$.
For comparison, we also plot the constraints from
SNe Ia uniformly distributed in the redshift bin
$0.5 < z < 1.8$ on the dark energy EOS parameter $w (0.5 < z < 1.8)$
in Fig.~\ref{fig:pw_SNIa_mz_uni}.
We can see that the contributions from GRBs are comparable to that
from SNe Ia.

\section{Summary}

We explored the GRBs' contributions in constraining the dark energy
EOS at high redshifts ($1.8 < z < 7$) and at middle redshifts
($0.5 < z < 1.8$). When constraining the dark energy EOS in a certain
redshift range, we allow the dark energy EOS parameter to vary only in
that redshift bin and fix EOS parameters elsewhere to $-1$.
We find that it is difficult to constrain the dark energy EOS
parameters beyond the redshifts of SNe Ia with GRBs unless some new
luminosity relations for GRBs with smaller scatters are discovered.
However, at middle redshifts, GRBs have contributions comparable with
SNe Ia in constraining the dark energy EOS.

\begin{acknowledgements}
This work was supported by the Scientific Research Foundation of the
Graduate School of Nanjing University (for Shi Qi),
the Jiangsu Project Innovation
for PhD Candidates CX07B-039z (for Fa-Yin Wang),
and the National Natural Science Foundation of China
under Grant No.~10473023.
\end{acknowledgements}

\bibliographystyle{aa}
\bibliography{dark_energy}

\end{document}